\documentclass[structabstract]{aa} 
\usepackage{graphicx}
\usepackage{txfonts}
\begin{document}
   \title{The low-mass stellar mass functions of rich, compact clusters in the Large Magellanic Cloud}


   \author{Q. Liu
      \inst{1,2}
   \and R. de Grijs
      \inst{3,1}
   \and L. C. Deng
      \inst{1}
   \and Y. Hu
      \inst{1,2}
   \and S. F. Beaulieu
      \inst{4}
}

   \institute{National Astronomical Observatories, Chinese Academy of Sciences,
             Beijing 100012, China\\
              \email{[liuq,licai,huyi]@bao.ac.cn}
         \and
             Graduate University of the Chinese Academy of Sciences, Beijing
100049, China
         \and
         Department of Physics \& Astronomy, The University of Sheffield,
Sheffield S3 7RH, UK\\
           \email{R.deGrijs@sheffield.ac.uk}
        \and
         D\'{e}partement de Physique, de G\'{e}nie Physique et d'Optique
and Centre de Recherche en Astrophysique du Qu\'ebec,\\ Universit\'{e}
Laval, Qu\'ebec, QC G1V 0A6, Canada
             }

   \date{}


\abstract
{We use \textsl{Hubble Space Telescope} photometry of six rich,
compact star clusters in the Large Magellanic Cloud (LMC), with ages
ranging from 0.01 to 1.0 Gyr, to derive the clusters' stellar mass
functions (MFs) at their half-mass radii.}
{The LMC is an ideal environment to study stellar MFs, because it
contains a large population of compact clusters at different
evolutionary stages. We aim to obtain constraints on the
\textit{initial} MFs (IMFs) of our sample clusters on the basis of
their present-day MFs, combined with our understanding of their
dynamical and photometric evolution.}
{We derive the clusters' present-day MFs below $1.0 M_{\odot}$ using
deep observations with the Space Telescope Imaging Spectrograph and
updated stellar population synthesis models.}
{Since the relaxation timescales of low-mass stars are very long,
dynamical evolution will not have affected the MFs below $1.0
M_{\odot}$ significantly, so that -- within the uncertainties -- the
derived MFs are consistent with the solar-neighbourhood IMF, at least
for the younger clusters.}
{The IMF in the low-density, low-metallicity environment of the LMC
disk is not significantly different from that in the solar
neighbourhood.}


   \maketitle
%

\section{Introduction}

The shape of the stellar initial mass function (IMF) is a very
important unresolved issue in modern astrophysics, because it plays a
crucial role in many of the remaining `big questions.' The IMF is
usually assumed to be universal, and best approximated by either a
power-law (e.g., Kroupa 2001) or a lognormal distribution (Chabrier
2003; Andersen et al. 2008). Kroupa (2001) studied the Galactic-field
IMF down to $0.01 M_{\odot}$ and derived a three-part power-law
function. Chiosi et al. (2007) obtained a similar mass function down
to $0.7 M_{\odot}$ based on their analysis of three clusters in the
Small Magellanic Cloud (SMC), and concluded that the IMF in SMC
clusters is in agreement with the `standard' Kroupa (2001)
solar-neighbourhood IMF. Da Rio et al. (2009) recently studied the
stellar association LH 95 in the Large Magellanic Cloud (LMC),
focusing on its pre-main-sequence (PMS) stars. Their results showed
that there are no significant differences between the IMFs of the
entire LH 95 region and those of three individual subclusters: they
all follow a multiple power-law distribution. Andersen et al. (2008)
studied the low-mass stellar mass distributions of seven star-forming
regions and concluded that the composite IMF is consistent with a
lognormal distribution. Paresce et al. (2000) obtained a similar
result after analysing the mass functions (MFs) of a dozen Galactic
globular clusters (GCs) for stellar masses below $1.0
M_{\odot}$. Thus, the form of the universal IMF appears to be best
approximated by both Kroupa (2001) broken power-law and lognormal
distributions (Chabrier 2003; Andersen et al. 2008; Covey et al. 2008;
Liu et al. 2009; Oliveira et al. 2009). To distinguish between either
shape, one would need to probe down to the stellar/brown-dwarf
transition region, which still poses a significant observational
challenge, particularly in (even the nearest) extragalactic
environments.

The evolution of the stellar MF is also important more generally,
because the IMFs of many clusters and galaxies cannot be observed
directly. However, we can derive their IMFs based on the present-day
MF (PDMF) if -- at least -- we understand its evolution in
detail. Star clusters, both open clusters and GCs, provide ideal
objects to tackle many astronomical problems, because all of their
member stars have approximately the same age and metallicity, and are
located roughly at the same distance. Although the GCs in the Milky
Way are relatively nearby and their members can be observed easily,
they are not well suited to study the evolution of the MF, because
Galactic GCs are all old (with ages $t \geq 10$ Gyr). They can
therefore only provide evolutionary information on long
timescales. Galactic open clusters, on the other hand, are only
effective tracers of MF evolution on short timescales, while they also
tend to be affected quite significantly by small-number
statistics. Ideally, therefore, we need rich massive clusters covering
a large age range to make significant progress on this important
problem. This makes the LMC an ideal laboratory, because it contains a
large population of rich star clusters with masses similar to Galactic
GCs and covering ages from 0.001 to 10 Gyr (e.g., Beaulieu et
al. 1999; Elson et al. 1999). This implies that we can study the MF at
almost all evolutionary stages using the rich, compact clusters in the
LMC. In the past, it proved impossible to resolve individual stars in
dense star clusters at the distance of the LMC ($\sim 50$ kpc), but
the unprecedented, high spatial resolution of the {\sl Hubble Space
Telescope (HST)} facilitates such studies today.

This is what we set out to do in this paper. In Section 2, we briefly
describe our {\sl HST} observations and give a basic overview of the
data-reduction procedures (see detailed steps in Liu et al. 2009). We
present and discuss our main results in Sections 3 and 4,
respectively, and provide a summary in Section 5.


\section{Observations and data reduction}
\label{sect:Obs and data}

\subsection{Observations and previous work}

As part of {\sl HST} programme GO-7307 we observed a carefully
selected cluster sample in the LMC, including six compact clusters in
three pairs (Pair I: NGC 1805 and NGC 1818, Pair II: NGC 1831 and NGC
1868, and Pair III: NGC 2209 and Hodge 14): see Table 1 for their
fundamental parameters. Our six sample clusters have ages of $10^{7} -
10^{9}$ yr, with the additional constraint that the two clusters in
each pair have similar ages, metallicities, total mass, and distance
from the LMC centre, yet different structural parameters (Beaulieu et
al. 1999).

\begin{table*}
\begin{centering}
 \caption[]{Fundamental parameters of our LMC cluster
sample.}\label{parameters}
\begin{tabular}{ccccccccccc}
  \hline\noalign{\smallskip}
Cluster &  log(age)$^{a}$  & $E(B-V)^{b}$ &$(m-M)_0^{b}$ & $\log(M_{\rm cl}/M_{\odot})^{c}$  & $R_{\rm core}^{c}$ & $D_{\rm LMC}^{d}$    \\
        &  [yr]           & (mag)     &   (mag)            & &
        (pc) & ($^{\circ}$) \\
  \hline\noalign{\smallskip}
NGC 1805  & 7.65$\pm$0.05   & 0.04   &18.59   &3.52$\pm$0.13  &
1.33$\pm$0.06   &
 3.86--4.00  \\ 
NGC 1818  & 7.65$\pm$0.05  & 0.03  & 18.58  & 4.13$^{+0.15}_{-0.14}$

& 2.45$\pm$0.09  & 3.47--3.61 \\
NGC 1831  & 8.75$\pm$0.05  & 0.00  & 18.58 & 4.81$\pm$0.13&
4.44$\pm$0.14& 4.82--4.85\\
NGC 1868  & 9.00$\pm$0.05 & 0.02 & 18.55 & 4.53$\pm$0.10 &
1.62$\pm$0.05 & 5.57--5.47\\
NGC 2209  & 9.10$\pm$0.05 & 0.07  & 18.39  & 5.03$^{+0.36}_{-0.6}$ &

5.43$\pm$0.33  & 5.48--5.43 \\
Hodge 14  & 9.30$\pm$0.05  & 0.04  & 18.49 & 4.33$^{+0.34}_{-0.28}$

& 1.80$\pm$0.14  & 4.19--4.37\\
  \noalign{\smallskip}\hline
\end{tabular}
\end{centering}
\flushleft {Notes: $D_{\rm LMC}$ is the distance from the LMC
centre; the two values indicate $D_{\rm LMC}$ to the optical,
geometrical centre (Bica et al. 1996) and the dynamical, rotation
centre (Westerlund 1990)}, respectively. References: $^a$ this paper
(the age uncertainties are driven by the discreteness of the Padova
isochrones), $^b$ Castro et al. (2001), $^c$ Mackey \& Gilmore
(2003), $^d$ Meurer, Cacciari, \& Freeman (1990).
\end{table*}

Much work has been done already based on these high-quality imaging
observations (e.g., Elson et al. 1998, 1999; Johnson et al. 2001;
Santiago et al. 2001; de Grijs et al. 2002a,b,c; Kerber et al. 2006;
Liu et al. 2009). de Grijs et al. (2002a,b) obtained the MFs of the
two youngest clusters in our sample for stellar masses above $1.0
M_{\odot}$, and concluded that they are largely similar to the
Salpeter (1955) IMF. Liu et al. (2009) analysed the MF of NGC 1818
below 1.0 M$_{\odot}$ and combined their low-mass MF with the
higher-mass results of de Grijs et al. (2002b). They found that the
IMF of NGC 1818 could be well approximated by both a Kroupa
(2001)-type broken power-law function and a lognormal distribution.

Our high-quality imaging observations were obtained with both the
Wide-Field and Planetary Camera-2 (WFPC2) and the Space Telescope
Imaging Spectrograph (STIS). WFPC2 is composed of four chips (each
containing $800\times800$ pixels), one Planetary Camera (PC) and three
Wide-Field (WF) arrays (WF2, WF3, and WF4). The pixel size of the PC
chip is 0.0455 arcsec (with a field of view of $\sim34\times34$
arcsec$^{2}$) and that of each WF chip is 0.097 arcsec (with fields of
view of $\sim150\times150$ arcsec$^{2}$). The STIS pixel size is
0.0507 arcsec and the corresponding field of view is about
$28\times52$ arcsec$^{2}$.

WFPC2 exposures through the F555W and F814W filters (roughly
corresponding to the Johnson-Cousins $V$ and $I$ bands, respectively)
were obtained with the PC chip centred on both the clusters' half-mass
radii (with a total exposure time of 2500 s in both filters; see for
more details Santiago et al. 2001; de Grijs et al. 2002a). Deep STIS
exposures in ACCUM imaging mode through the F28$\times$50LP long-pass
filter (central wavelength $\simeq 7230$\AA) were also obtained,
centred on the clusters' half-mass radii (with total exposure times of
2950 s for Pairs I and III, and 2890 s for Pair II; see Elson et
al. 1999). To obtain `clean' MFs, we must subtract the background
stellar contribution. We therefore also obtained deep WFPC2 images
from the {\sl HST} Data Archive of the general LMC background through
the F555W and F814W filters, with exposure times of 7800 and 5200 s,
respectively (see Castro et al. 2001; Santiago et al. 2001; de Grijs
et al. 2002a; Liu et al. 2009).

\subsection{Data reduction and photometry}

We use the same method for data reduction and photometry as in Liu et
al. (2009). Aperture photometry\footnote{See Liu et al. (2009) for a
discussion of the pros and cons of using aperture photometry versus
point-spread function fitting.} was performed on our WFPC2 and STIS
images using the {\sc iraf/apphot}\footnote{The Image Reduction and
Analysis Facility ({\sc iraf}) is distributed by the National Optical
Astronomy Observatories, which is operated by the Association of
Universities for Research in Astronomy, Inc., under cooperative
agreement with the US National Science Foundation.} package, and
2-pixel apertures were adopted since this produced the smallest
photometric errors and the tightest cluster main sequences.

\begin{figure}
\centering
\includegraphics[width=9.2cm]{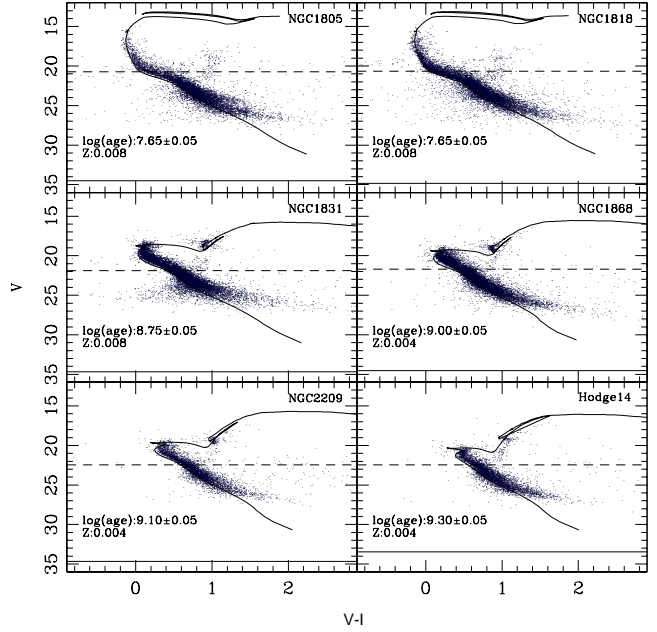}
\caption{Colour-magnitude diagrams of our LMC clusters and the
best-fitting Padova isochrones (Girardi et al. 2000). The horizontal
dashed and solid lines in each panel represent the upper and lower
magnitude limits, respectively, of the parameter space covered by our
STIS observations.} \label{age}
\end{figure}

\begin{figure}
\centering
\includegraphics[width=9.2cm]{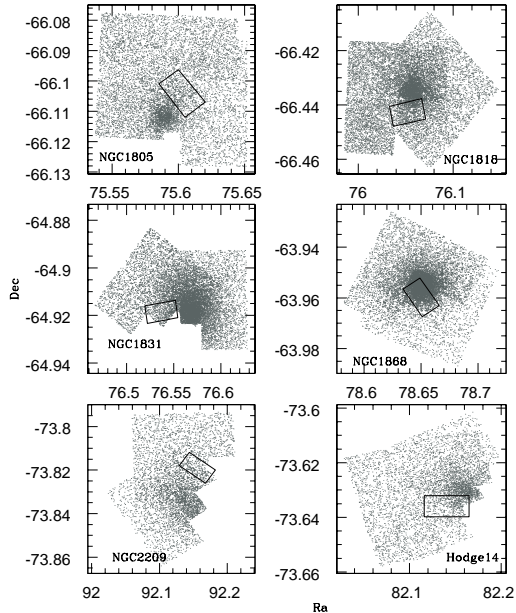}
\caption{Spatial distribution of the stars in our LMC clusters. The
dots represent the stars detected in the WFPC2 observations, while the
solid rectangular shapes indicate the areas covered by our STIS
observations, which are used for further analysis in this paper.}
\label{space}
\end{figure}

We used the relations of Whitmore et al. (1999) to correct the
resulting photometry for the time-dependent charge-transfer efficiency
effects and {\sc iraf/stsdas}\footnote{{\sc stsdas}, the Space
Telescope Science Data Analysis System, contains tasks complementary
to the existing {\sc iraf} tasks. Version 3.1 was adopted for the data
reduction performed in this paper.} to rectify the geometric
distortions of the WFPC2 chips. We subsequently applied aperture
corrections based on the model PSFs generated by TinyTim (Krist \&
Hook 2001). Finally, we used the relations of Holtzman et al. (1995)
to convert the aperture-corrected F555W and F814W magnitudes to the
Johnson-Cousins $V$ and $I$ passbands. Figures \ref{age} and
\ref{space} show, respectively, the resulting WFPC2-based
colour-magnitude diagrams (CMDs) of all sample clusters and the
spatial distributions of the stars in our LMC cluster fields as
observed with both WFPC2 and STIS (dots and rectangular outlines,
respectively).

\subsection{Completeness corrections and background subtraction}

One of the most difficult problems we faced to derive clean MFs
involved correcting for sampling incompleteness, which is usually a
function of position in a cluster. We used the same method as in Liu
et al. (2009), in essence a slightly modified version of the approach
used in de Grijs et al. (2002a), who computed these corrections in
circular annuli around the cluster centres. We computed the
completeness corrections for the entire STIS chip, because the STIS
observations were centred on the low-density half-mass radius and the
effects of sampling incompleteness are constant across our STIS field
within the observational uncertainties (e.g., de Grijs et al. 2002a;
Liu et al. 2009). We used the same method to compute the equivalent
completeness corrections for the background field. We added an
area-dependent number of artificial sources of Gaussian shape to each
STIS exposure with input magnitudes between 16.0 and 30.0 mag, in
steps of 0.5 mag. We then adopted the same photometric analysis method
for the fields including both the `real' cluster (and background)
stars and the artificial stars, to asses how many artificial stars
could be detected after correction for blends and superpositions. For
the analysis in this paper, we only consider magnitude ranges that are
$\geq 50$\% complete.

To obtain a clean MF, we must subtract the contamination by the
background field, because its stellar mass distribution is generally
different from that of the clusters (cf. Castro et al. 2001). We do
not have a background field in the STIS F28$\times$50LP passband.
Instead, we used a general WFPC2/F814W background-field observation.
However, because our STIS observations are much deeper than the WFPC2
data, we could not directly subtract the full background effects from
the observations. Gouliermis et al. (2006a) suggested that the stellar
mass distribution in the LMC disk follows a broken power-law
distribution. We therefore adopted a power law (with slope
$\Gamma=1.87$, where the IMF, $\xi(m_\ast) \propto m^{\Gamma}$) to
approximate and extrapolate the stellar mass distribution in the
general LMC background down to 0.1 $M_{\odot}$ (see Liu et al. 2009,
their fig. 5).

\begin{figure}
\centering
\includegraphics[width=\columnwidth]{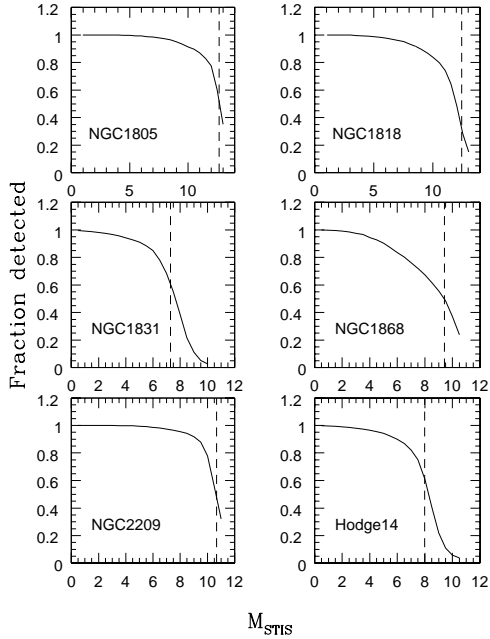}
\caption{Completeness ratios of all sample clusters. The dashed
lines represent the 50\% completeness limits for the individual
clusters.} \label{allcomplete}
\end{figure}

\begin{figure}
\centering
\includegraphics[width=\columnwidth]{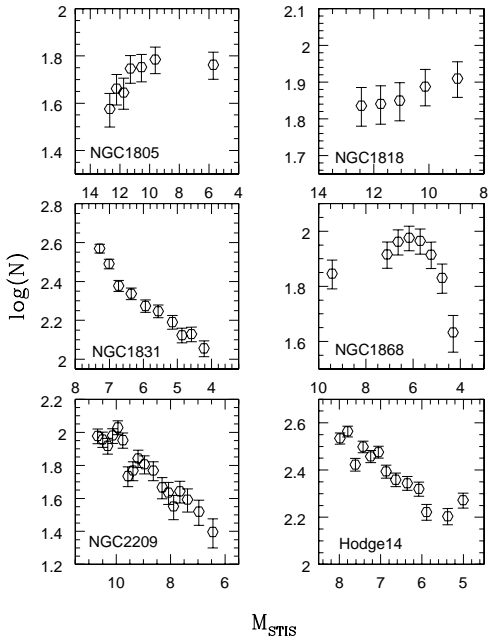}
\caption{Completeness-corrected, background-subtracted LFs of all
sample clusters, for completeness fractions $\ge 50$\%.}
\label{allLFs}
\end{figure}

In Figs. \ref{allcomplete} and \ref{allLFs} we show the photometric
completeness fractions and the resulting luminosity functions (LFs)
for all of our sample clusters, respectively. The LFs have been fully
completeness corrected and background subtracted. We only include
magnitude bins for which the observational completeness fractions are
greater than 50\%.

\section{Analysis}
\label{sect:analysis}

\subsection{Age, metallicity, and evolutionary models}

Santiago et al. (2001), de Grijs et al. (2002b,c), and Kerber et al.
(2006) studied the MFs of our sample clusters above $1.0 M_{\odot}$,
based on WFPC2 data. Liu et al. (2009) studied the MF of NGC 1818
below 1.0 $M_{\odot}$ based on STIS data. Because the STIS data are
much deeper than the WFPC2 measurements (e.g., de Grijs et al.
2002a), here we study the MFs of all six LMC clusters from {\sl HST}
programme GO-7307 based on the STIS observations.

The ages of our six sample clusters cover a large range, from $10^{7}
- 10^{9}$ yr. Most of the stars below $1.0 M_{\odot}$ of the clusters
in the youngest of pair, NGC 1805 and NGC 1818, are still on the PMS
(Liu et al. 2009). The evolution of PMS stars is still uncertain
(Baraffe et al. 1997, 1998). White et al. (1999) concluded that the
models of Baraffe et al. (1997, 1998) resulted in the most consistent
ages and masses, on the basis of a comparison of six PMS evolutionary
model sets (Park et al. 2000).  Because our STIS observations were
obtained in a single passband only, we cannot derive the PMS ages and
metallicities on the basis of CMD analysis. However, the cluster stars
from our WFPC2 and STIS observations occupy a common locus on the CMD,
so we can obtain the basic cluster parameters by adopting Padova
isochrones (Girardi et al. 2000) to fit the faint end of the main
sequence, and the Baraffe et al. (1997, 1998) models for the sequence
of low-mass stars, i.e., by extrapolating the WFPC2 observations to
fainter luminosities (see, for details, Liu et al. 2009).

The other sample clusters are old enough for all low-mass stars to
have evolved onto the main sequence, so that we can get their ages and
metallicities using Padova isochrones for main-sequence fitting.  The
best fits and the resulting basic cluster parameters are shown in
Fig. \ref{age}. We used the Padova isochrones to fit the main-sequence
ridge line in the CMD with different metallicities and ages for each
cluster. Next, we compared the quality of all fits and adopted the
best-fitting metallicity and age for the cluster of interest. The
small spread of the main-sequence data points will cause a small
uncertainty in the age determination for each cluster, which we
characterise by using the discreteness of the Padova isochrones (in
steps of 0.05 dex in age) as a proxy for the age uncertainties. In
fact, for each cluster the nearest isochrones to the best-fitting
model provide markedly worse fits to the main-sequence turn-off
location, so that we are confident that our fits are robust.

In Liu et al. (2009) we added the F28$\times$50LP filter to the
Baraffe et al. (1998) model suite. These models cover a range of
metallicities, enabling us to choose the most relevant mass-luminosity
relation for conversion of the F28$\times$50LP magnitudes to
individual stellar masses.

A significant body of work exists in the literature to support our
choice of metallicity for the individual clusters. For NGC 1805,
Johnson et al. (2001) used {\sl HST} CMDs to derive a mean [Fe/H]
$\sim 0$ (solar metallicity), although Meliani et al. (1994) had
argued previously that the cluster's most appropriate metallicity was
$Z = 0.008$ (where $Z_\odot = 0.020$), i.e., the average metallicity
of the young LMC field population. On this basis, we adopted $Z =
0.008$ for this cluster, corresponding to [Fe/H] $\simeq -0.37$
(assuming a one-to-one correlation between metallicity and iron
abundance). Using solar metallicity instead, we derive an age for NGC
1805 of $\log(t/{\rm yr}) \simeq 7.50\pm0.05$ using Padova
isochrones. While the slopes of the resulting MFs based on either
metallicity are similar (see Fig. \ref{MF.PairI}, top), the
calibration of the solar-metallicity MFs would shift to higher masses.

NGC 1818 has been the subject of a large number of studies aimed at
determining its metallicity. Although Johnson et al. (2001) obtained
solar metallicity from {\sl HST} CMD analysis, most other modern
(predominantly spectroscopic) determinations centre around either
[Fe/H] $= -0.37 \pm 0,03$ (e.g., Jasniewicz \& Th\'evenin 1994;
Bonatto et al. 1995) or [Fe/H] $\sim -0.8$, roughly corresponding to
$Z=0.003$ (e.g., Meliani et al.  1994; Will et al. 1995; Oliva \&
Origlia 1998). The MF slopes based on a metallicity of $Z=0.008$ are
steeper than for $Z=0.004$ (the lowest metallicity isochrone available
for the cluster's young age\footnote{Although the Padova models
include isochrones for $Z=0.001$, they are only provided for ages in
excess of $\log(t/{\rm yr})=7.80$. Use of the $Z=0.004$ isochrone
results in a cluster age of $\log(t/{\rm yr})=7.65\pm 0.10$. For the
low-mass PMS stars, we use $Z = 0.003$ and $\log(t/{\rm yr}) = 7.25$
based on the Baraffe et al.  (1997, 1998) models.}) (see Fig.
\ref{MF.PairI}, bottom), although they are both in good agreement with
the Kroupa (2001) IMF slope $(\Gamma=0.3\pm0.5)$.

The most recent metallicity estimates for NGC 1831 converge to [Fe/H]
$\sim -0.35$ (Bonatto et al. 1995, based on UV spectroscopy), while
Vallenari et al. (1992) similarly suggested a best estimate of $Z =
0.008$ based on their analysis of the literature on this cluster at
the time of their publication, but see Mateo et al. (1987) and
Olszewski et al. (1988, 1991) for close-to-solar abundance
estimates. Given the current observational status for this cluster, we
adopted $Z = 0.008$ as a compromise.

Both NGC 1868 and Hodge 14 are somewhat more metal-poor than the
younger clusters in our sample. For NGC 1868, Bica et al. (1986)
reported [Fe/H] $= -0.6 \pm 0.35$, consistent with Olszewski et al.'s
(1991) spectroscopic metallicity determination, [Fe/H] $= -0.50$.
Similarly, Jensen et al. (1988) and Olszewski et al. (1991) used
spectroscopy of Hodge 14 to obtain [Fe/H] $= -0.66 \pm 0.2$. For both
clusters we adopted $Z = 0.004$, corresponding to [Fe/H] $= -0.68$.

Finally, NGC 2209 is the lowest-metallicity cluster in our sample,
with [Fe/H] $= -0.9 \pm 0.3$ (Bica et al. 1986; see also Chiosi et
al. 1986; Dottori et al. 1987; Frogel et al. 1990). Based on a careful
analysis of the goodness-of-fit parameters for the range of
metallicities provides by the Padova isochrones, we adopted $Z =
0.004$ ([Fe/H] $= -0.68$) for NGC 2209; its CMD is much more poorly
approximated for [Fe/H] $=-0.9$.

Metallicity does not play an important role in our analysis of the MF
{\it slopes}. The models of Baraffe et al. (1997, 1998) for different
metallicities yield similar main sequences at low mass.

\subsection{Mass functions}

We adopted the same method as Liu et al. (2009) to derive the MFs of
the clusters of pair I. Although the age of NGC 1805 and NGC 1818 is
$\log(t/{\rm yr}) \simeq 7.65$, many low-mass stars are still on the
PMS. The ages of the PMS stars in both clusters are about $\log(t/{\rm
yr}) \simeq 7.25 \pm 0.40$ (Liu et al. 2009). All stars in the
clusters of pairs II and III have already evolved onto the main
sequence.

\begin{figure}
\centering
\includegraphics[width=0.495\textwidth, angle=0]{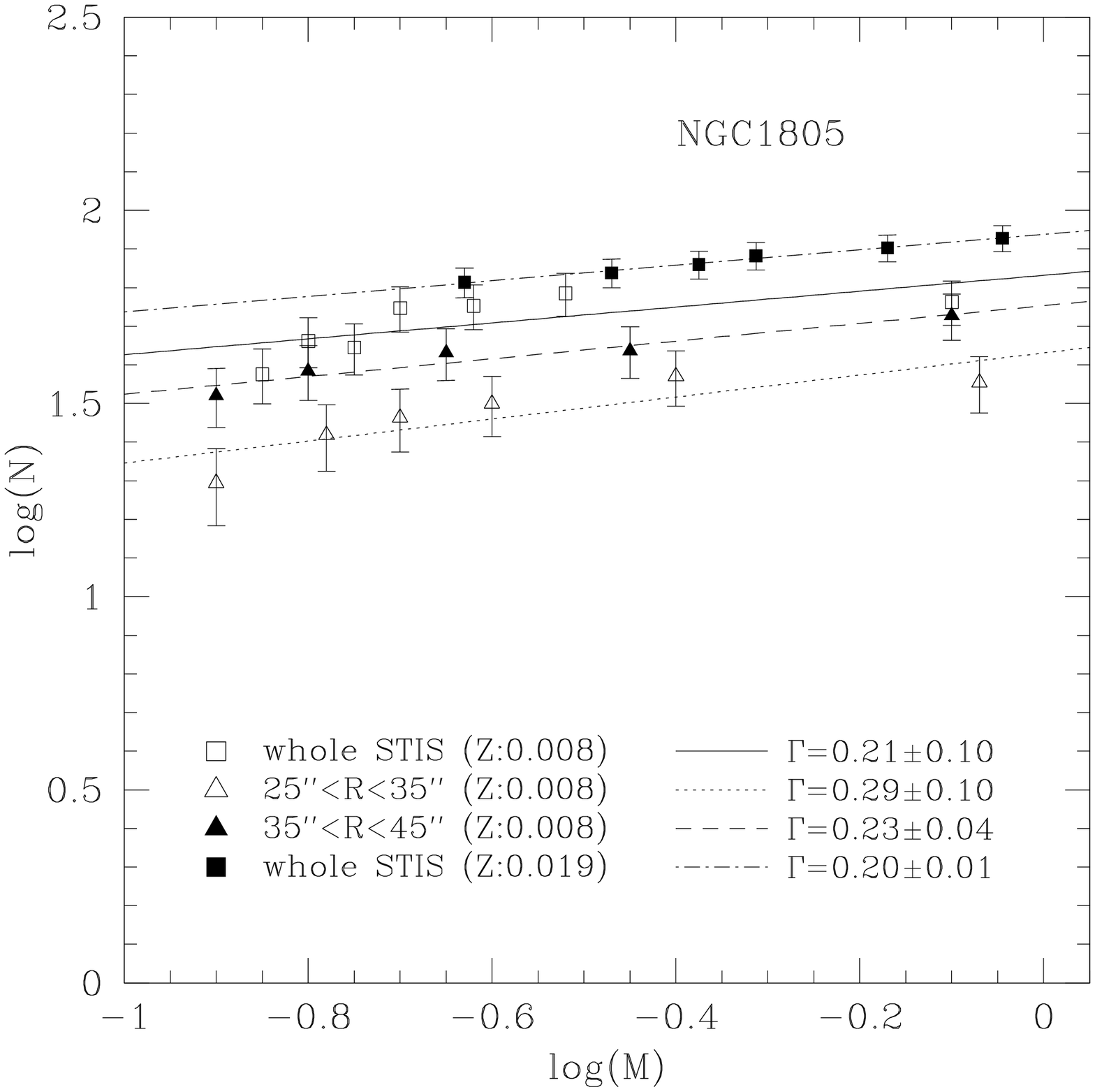}
\includegraphics[width=0.495\textwidth, angle=0]{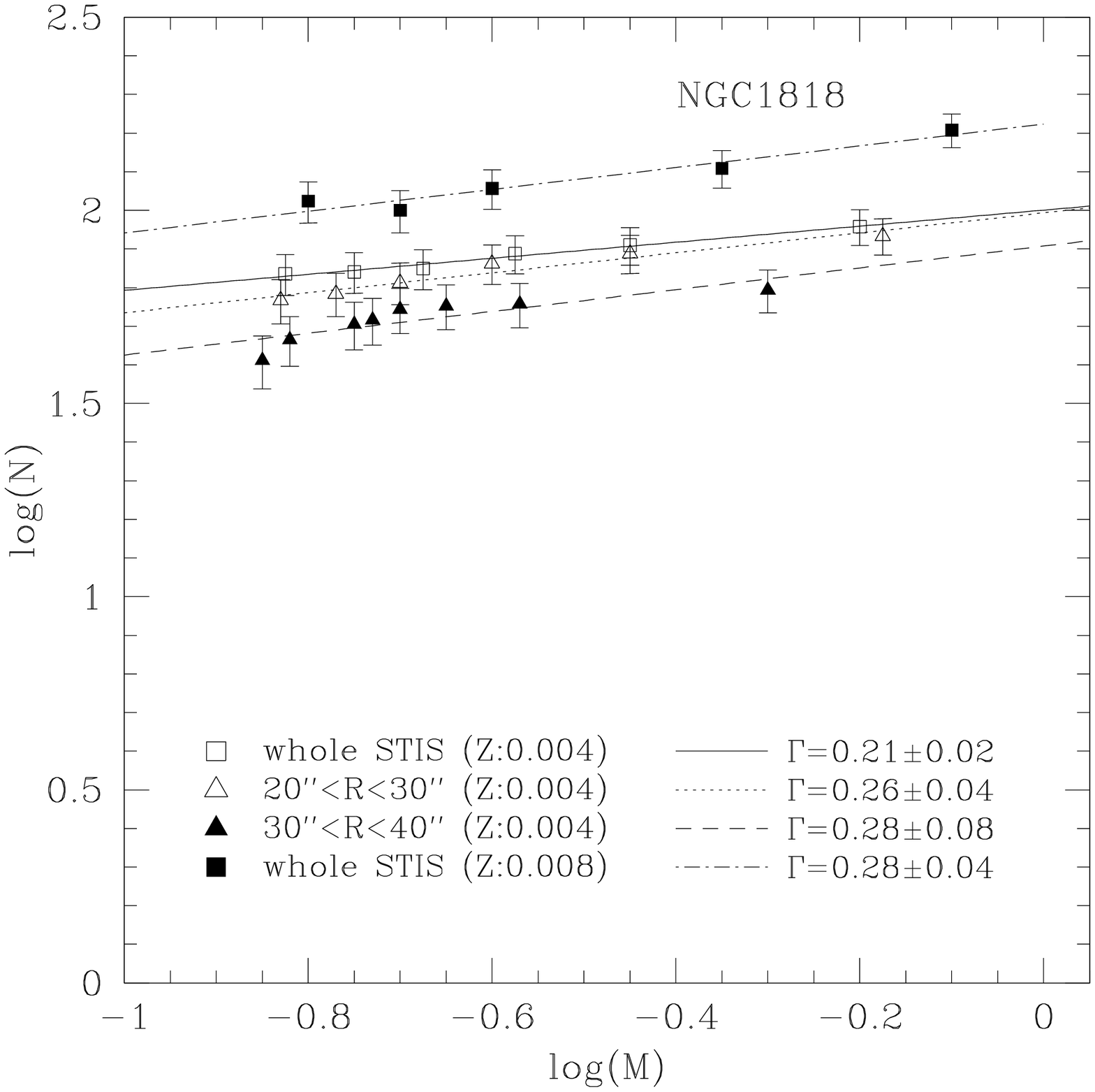}
\caption{Low-mass cluster MFs for Pair I, NGC 1805 and NGC 1818.}
\label{MF.PairI}
\end{figure}

\begin{figure}
\centering
\includegraphics[width=0.495\textwidth, angle=0]{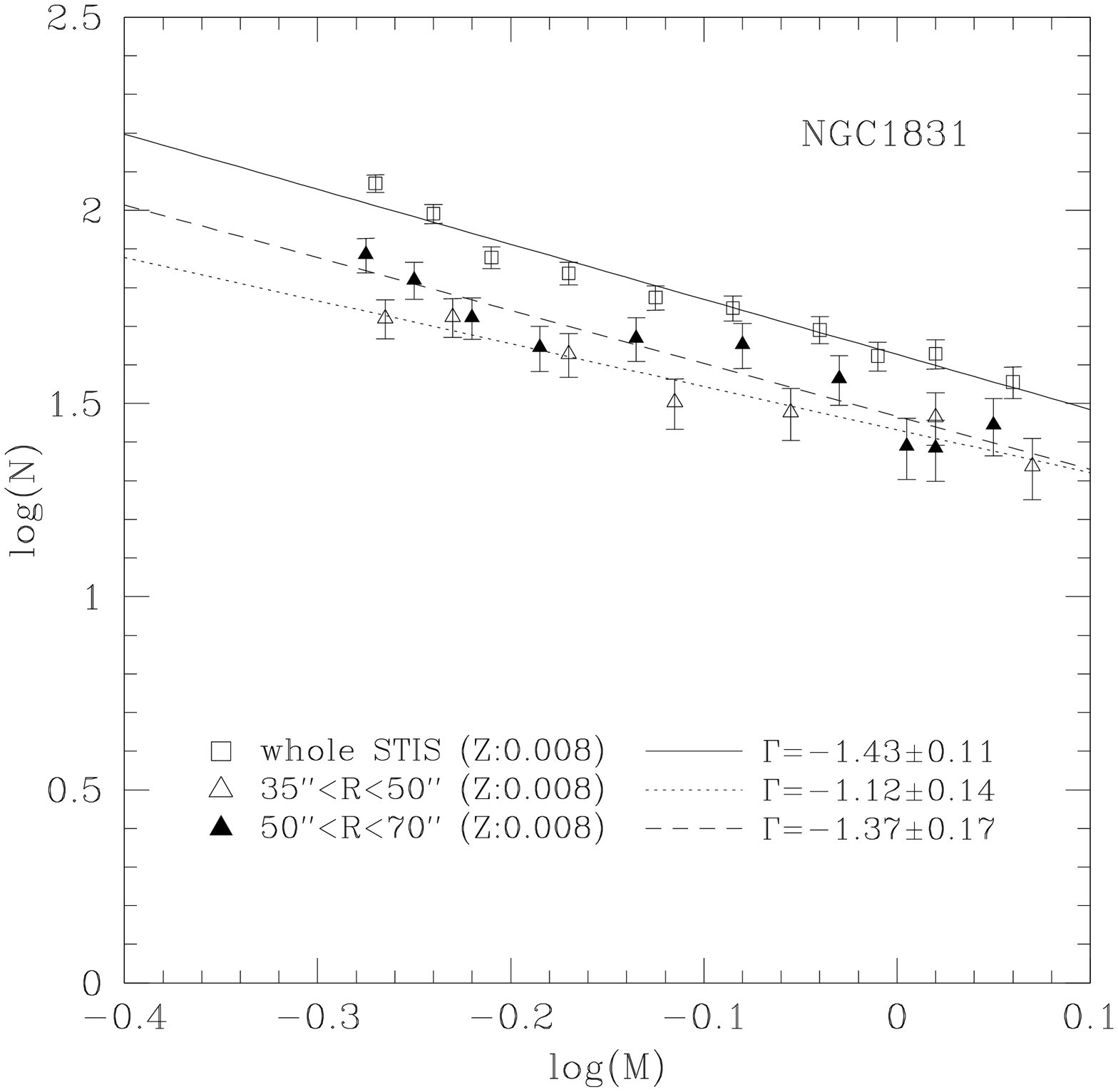}
\includegraphics[width=0.495\textwidth, angle=0]{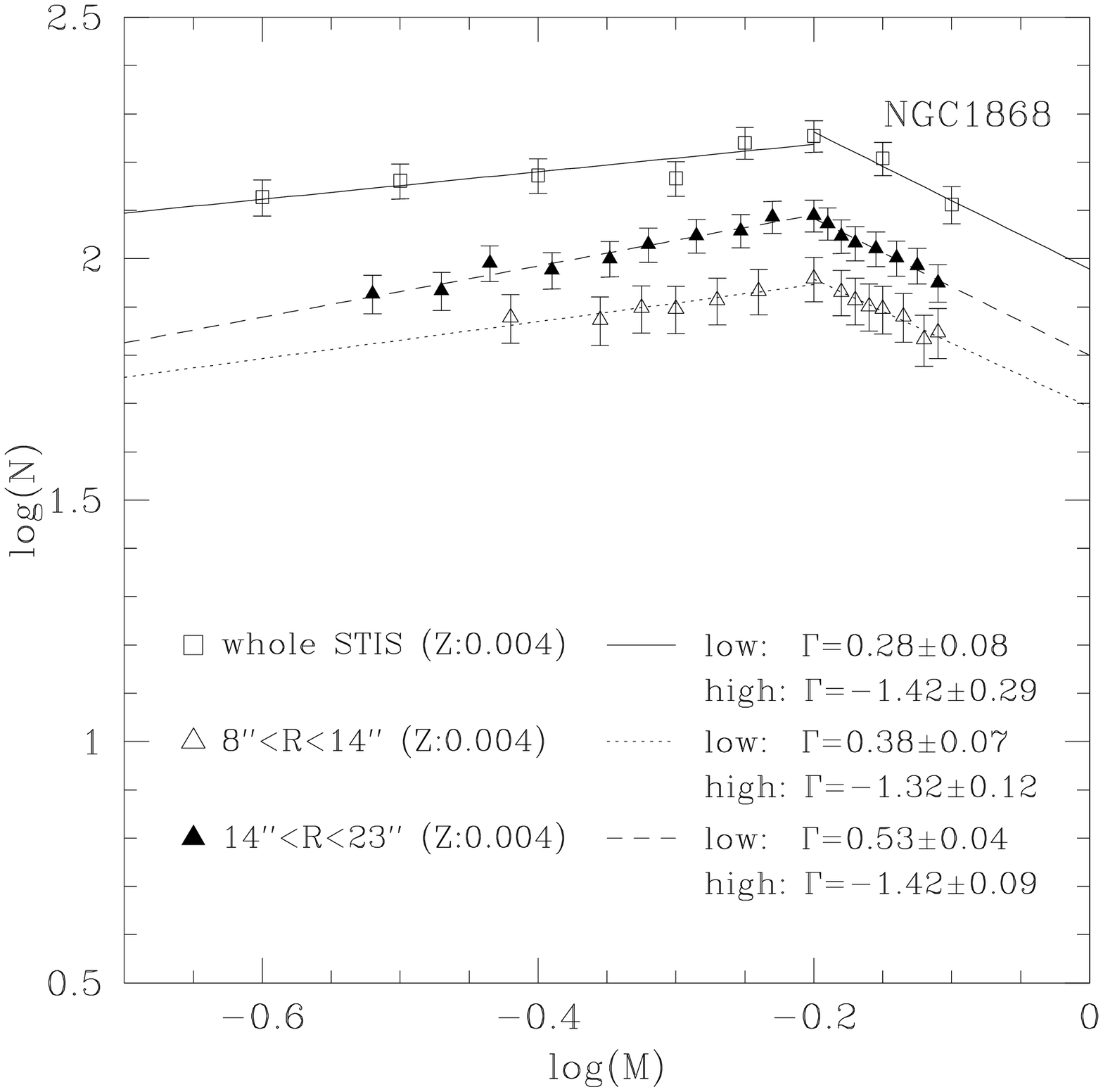}
\caption{Low-mass cluster MFs for Pair II, NGC 1831 and NGC 1868.}
\label{MF.PairII}
\end{figure}

\begin{figure}
\centering
\includegraphics[width=0.495\textwidth, angle=0]{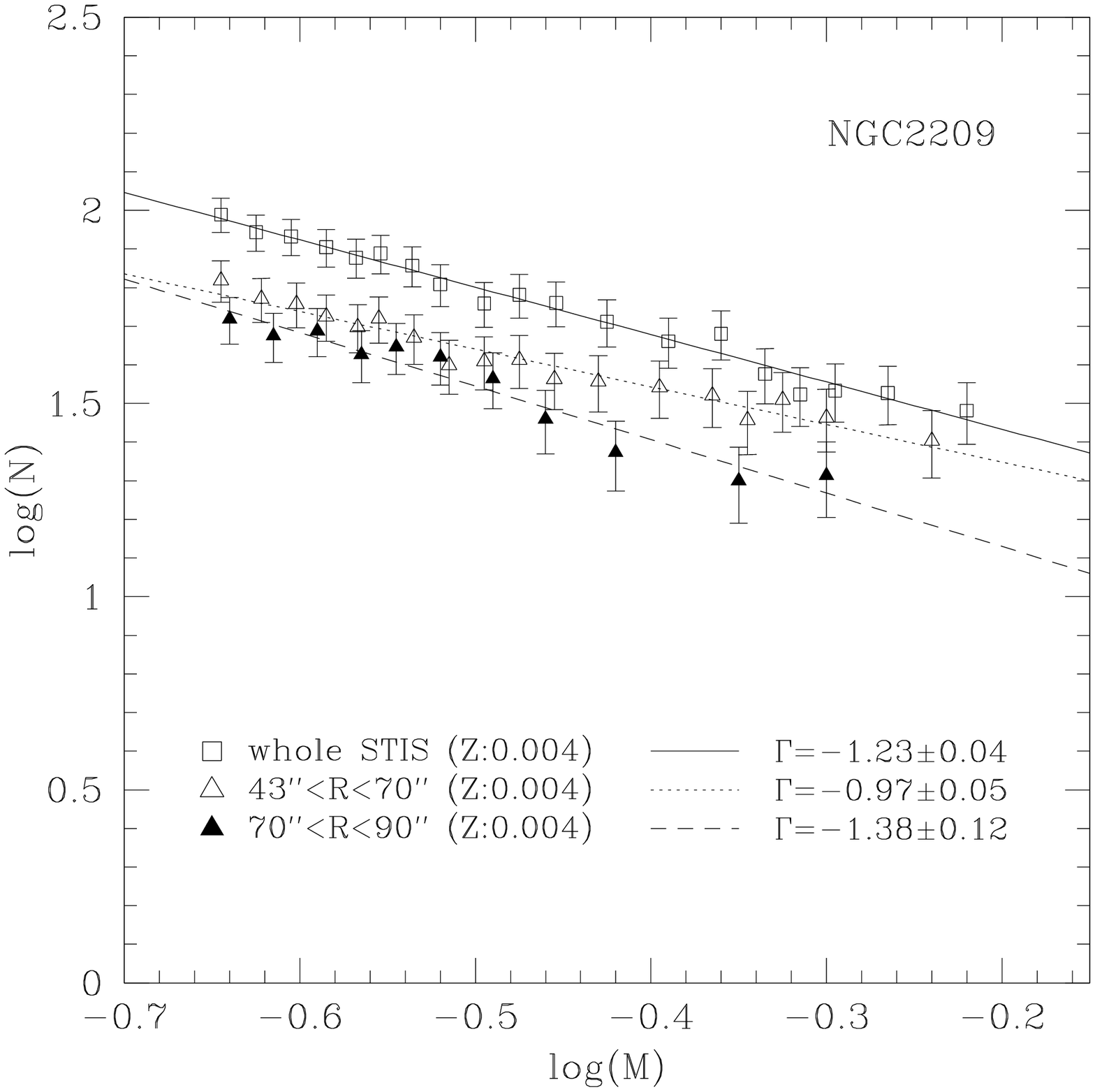}
\includegraphics[width=0.495\textwidth, angle=0]{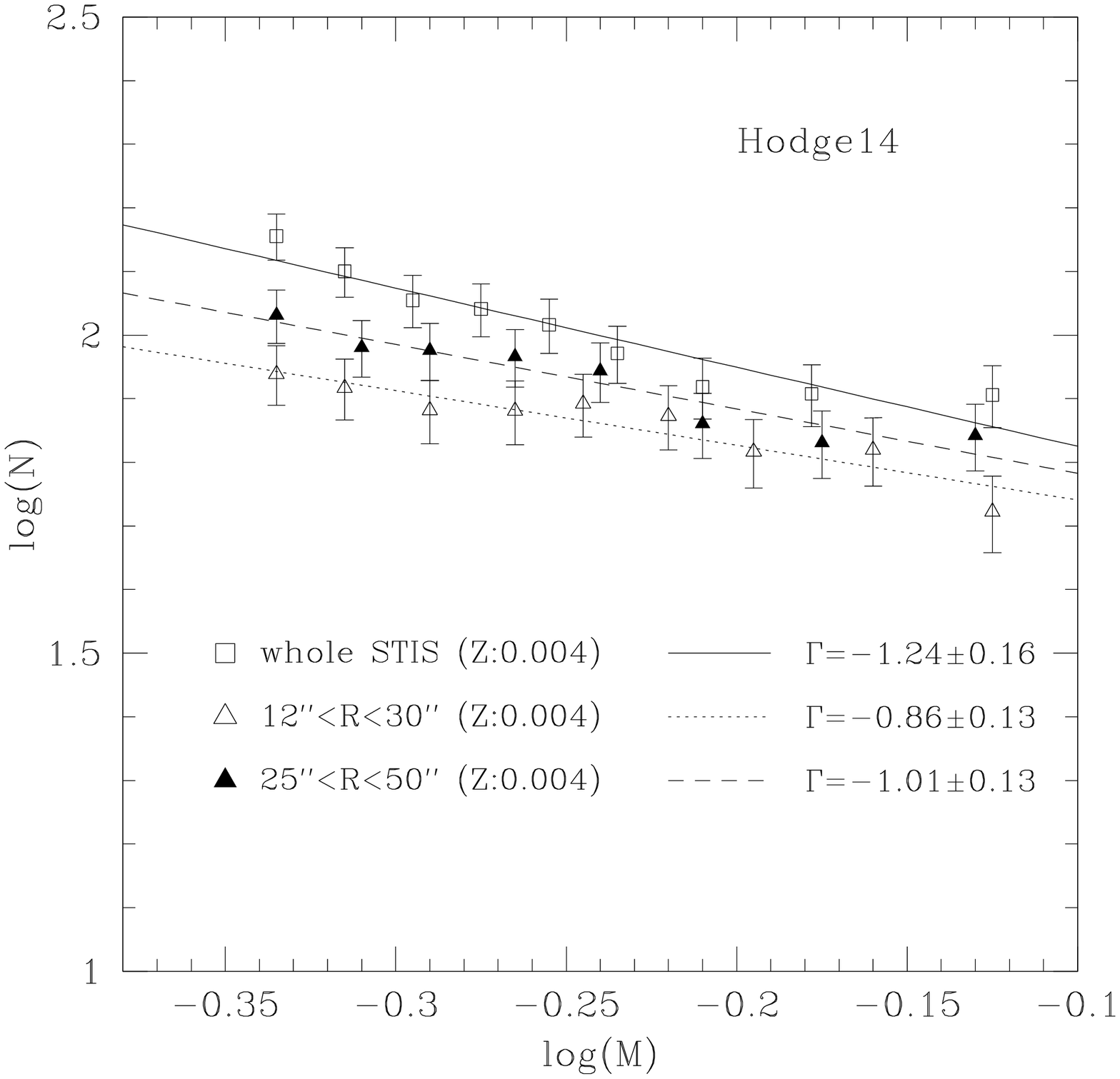}
\caption{Low-mass cluster MFs for Pair III, NGC 2209 and Hodge 14.}
\label{MF.PairIII}
\end{figure}

Several studies explored the effects of mass segregation in these
clusters (based on the same WFPC2 observations used here) by dividing
the full field of view into a number of smaller areas at a range of
distances from the cluster centres (e.g., Santiago et al.  2001; de
Grijs et al. 2002a,b,c; Kerber et al. 2006). However, our STIS
observations were taken at the half-mass radii of our sample clusters
and the STIS field is much smaller than that of the combined set of
WFPC2 observations, so we limited our analysis to both the entire STIS
region and areas at two different radii.  Figures \ref{MF.PairI},
\ref{MF.PairII}, and \ref{MF.PairIII} show the cluster MFs of (and
best fits to) different pairs for stellar masses below $1.0
M_{\odot}$, both for the full STIS field and for areas limited by
radial range. The results have been corrected for sample
incompleteness and background contamination. In Liu et al. (2009) we
combined the NGC 1818 MFs from WFPC2 and STIS. Using WFPC2, we
detected more stars than on the basis of our STIS data for some mass
ranges, so the slope of the MF in Liu et al. (2009),
$\Gamma=0.46\pm0.10$, is steeper than the slope obtained here
($\Gamma=0.28\pm0.04$). However, they are both in agreement with that
of the Kroupa (2001) IMF, $\Gamma=0.3\pm0.5$, within the
uncertainties.

The MFs of NGC 1805, NGC 1818, and NGC 1868 show the same trend, as
do the MFs of NGC 1831, NGC 2209, and Hodge 14. Figure
\ref{MF.PairII} shows the change in trend most clearly, for our
intermediate-age cluster pair. Because NGC 1868 is much more compact
than NGC 1831, its {\it dynamical} age is much older. This implies
that it will be much more evolved dynamically than NGC 1831, hence
resulting in a turnover at much lower masses (outside of our
observational range). Similarly, NGC 2209 and Hodge 14 are an order
of magnitude older (from both a stellar evolution and a dynamical
point of view) than the Pair II clusters, hence exhibiting declining
MFs in the mass range of interest here (see Fig. \ref{MF.PairIII}).

\begin{figure}
\centering
\includegraphics[width=0.42\textwidth, angle=0]{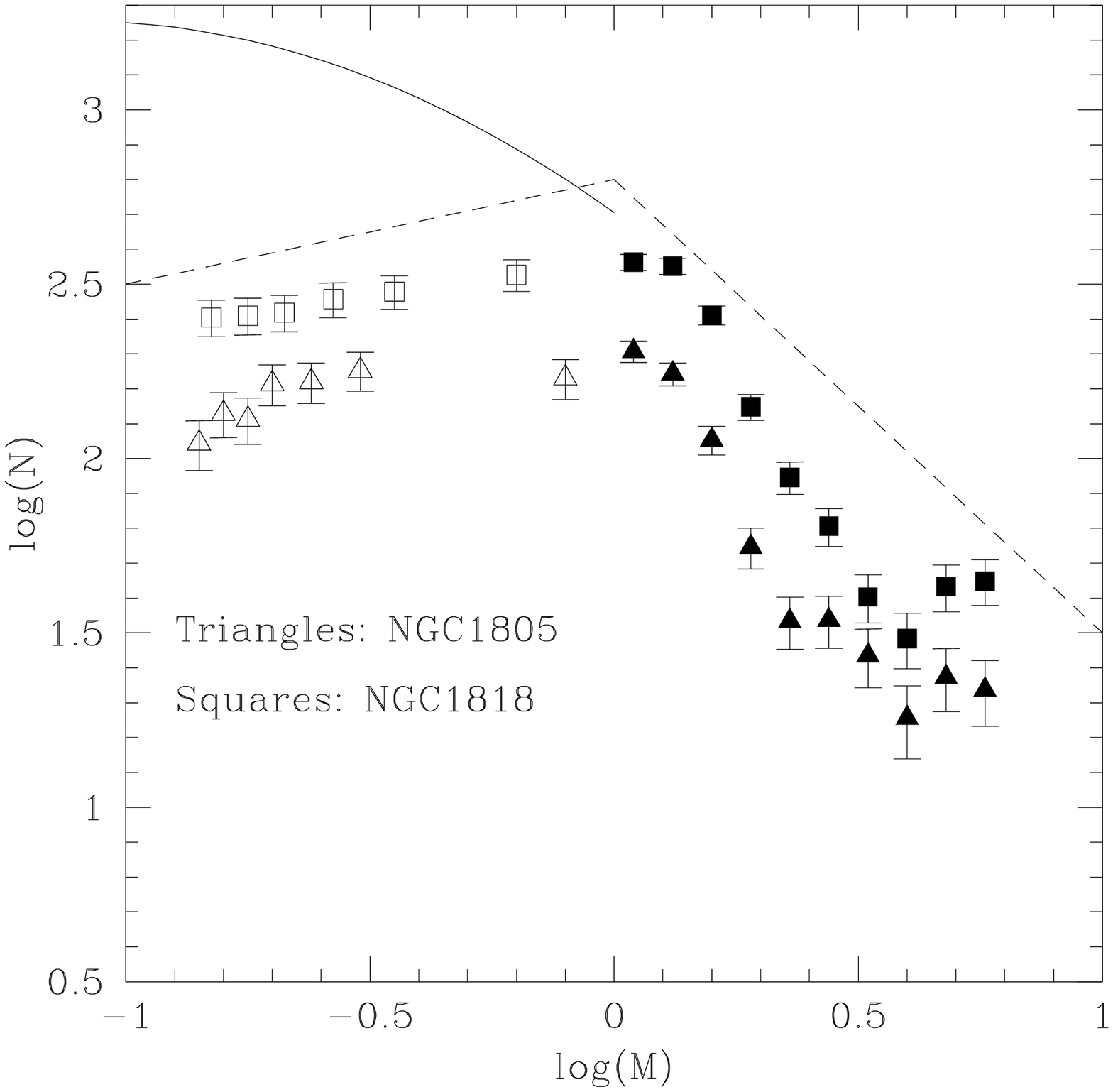}
\includegraphics[width=0.42\textwidth, angle=0]{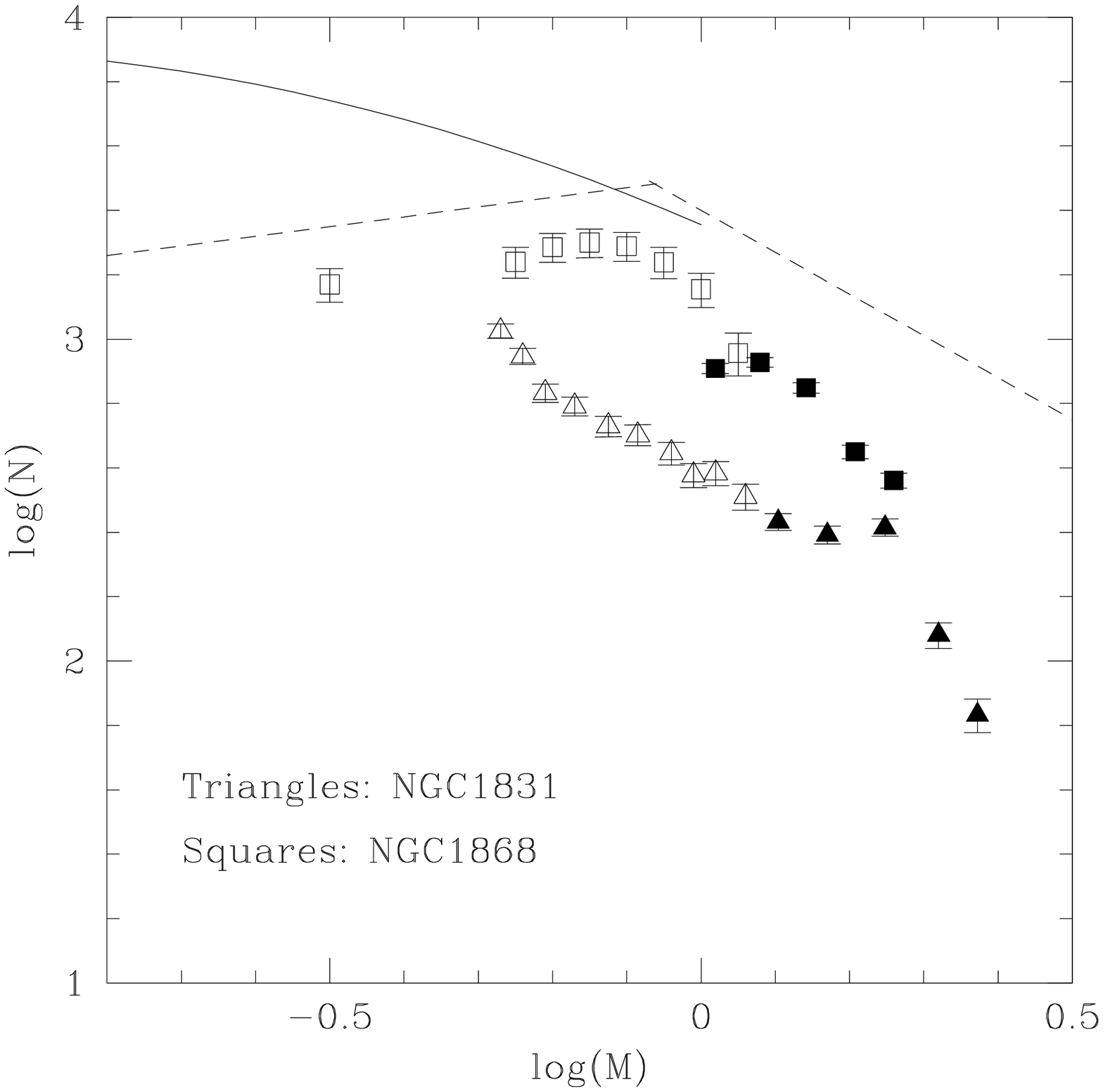}
\includegraphics[width=0.42\textwidth, angle=0]{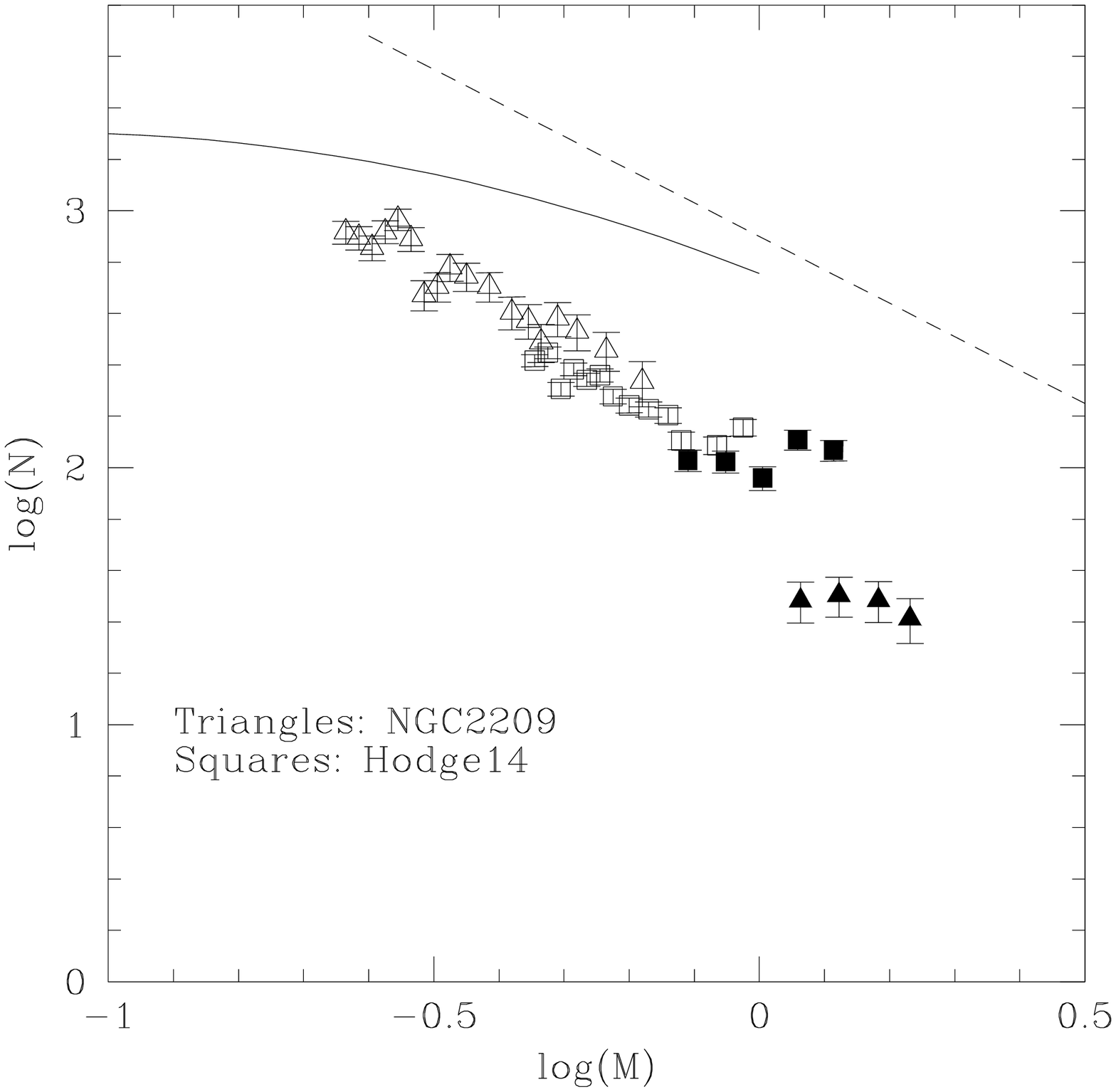}
\caption{Full MFs of the sample clusters arranged by pair: all open
points are from this paper, based on STIS observations, while the
solid points are from de Grijs et al. (2002c), based on WFPC2
observations. The dashed line represents a broken power-law IMF,
adopting the Kroupa (2001) high- and low-mass slopes, and the solid
curve shows the original Chabrier (2003) lognormal distribution,
defined below $1.0 M_{\odot}$, but offset from the data points for
reasons of clarity.} \label{LMCMF}
\end{figure}

We obtained all cluster MFs based on only single stars, neglecting
unresolved binary and other multiple stars. As discussed in Liu et al.
(2009), Kerber et al. (2006) analysed the effect of binarity on
cluster MFs. They found that MFs with binary fractions of unity and
0\% are identical within the observational uncertainties (cf. Liu et
al. 2009 for NGC 1818), so that we can justifiably ignore the effects
of binarity in the context of the low-mass {\it system} MFs derived in
this paper (and in Liu et al. 2009).

\section{Discussion}
\label{sect:discussion}

\begin{table}
\centering
\begin{minipage}{140mm}
\caption{Mass functions.}\label{number.MF}
\begin{tabular}{@{}ccccc@{}}
\hline
Cluster&$\log(m_\ast/M_\odot$) & $N$ & $\log(m_\ast/M_\odot$) & $N$ \\
\hline
NGC 1805&$-$0.85& 110 & 0.20 & 113\\
&$-$0.80& 135 & 0.28 & 56\\
&$-$0.75& 129 & 0.36 & 34\\
&$-$0.70& 164 & 0.44 & 34\\
&$-$0.62& 166 & 0.52 & 27\\
&$-$0.52& 179 & 0.60 & 18\\
&$-$0.10& 170 & 0.68 & 24\\
&   0.04& 203 & 0.76 & 22\\
&   0.12& 175 \\
\hline
NGC 1818&$-$0.825& 254& 0.20 & 258 \\
&$-$0.75& 257 & 0.28 & 141\\
&$-$0.675& 262& 0.36 & 88 \\
&$-$0.575& 286 & 0.44 & 64\\
&$-$0.45& 301 & 0.52 & 40\\
&$-$0.20 & 336& 0.60 & 30 \\
&  0.04 & 366 & 0.68 & 43\\
&  0.12 & 357 &0.76&45\\
\hline
NGC 1831&$-$0.27& 1059&  0.02 & 383\\
&$-$0.24& 884 & 0.06&324\\
&$-$0.21& 681 &0.104 & 271\\
&$-$0.17 &620 &0.17 & 247\\
&$-$0.125 & 536 & 0.248 & 260\\
&$-$0.085&  504 & 0.32 & 121\\
&$-$0.04& 442 & 0.372 & 68\\
&$-$0.01& 378\\
\hline
NGC 1868&$-$0.50& 1479 &0.00 & 1429\\
&$-$0.25& 1738& 0.05 & 906 \\
&$-$0.20& 1932 &0.08 & 847 \\
&$-$0.15& 1995 &0.14 & 706\\
&$-$0.10 &1945 & 0.21 & 447\\
&$-$0.05 & 1734 & 0.26 & 364\\
\hline
NGC 2209&$-$0.635& 825 &$-$0.38& 403 \\
&$-$0.615& 786 &$-$0.355 & 374\\
&$-$0.595& 718 &$-$0.335 &309\\
&$-$0.575& 828 &$-$0.31 & 381 \\
&$-$0.555 &927 &$-$0.28 & 340\\
&$-$0.535 &777 &$-$0.235 & 287\\
&$-$0.515 &471 &$-$0.18 & 216\\
&$-$0.495& 508 & 0.063& 30\\
&$-$0.475& 603  & 0.122 & 32\\
&$-$0.45&  555 &0.183 & 31\\
&$-$0.415 &508 &0.23 & 26\\
\hline
Hodge 14&$-$0.345& 262  &$-$0.17& 169\\
&$-$0.325& 281 &$-$0.14& 160\\
&$-$0.305& 203 &$-$0.12 & 128\\
&$-$0.285& 242  &$-$0.065 &122\\
&$-$0.265 &220 &$-$0.025 & 143\\
&$-$0.245 &229 &0.005 & 91\\
&$-$0.225 &190 &0.059 & 129\\
&$-$0.20& 175 &0.114 & 117\\
\hline
\end{tabular}
\end{minipage}
\end{table}

Owing to both the large distances involved and observational
limitations, it is difficult to obtain deep stellar MFs in
extragalactic environments. Kroupa (2001) studied the
solar-neighbourhood IMF down to $\sim 0.01 M_{\odot}$ and reported his
often-quoted broken power-law distribution. Paresce et al. (2000)
analysed the MFs of a dozen Galactic GCs down to $0.1 M_{\odot}$ and
derived a lognormal distribution below $1.0 M_{\odot}$. Much effort
has also focussed on studying stellar MFs in the LMC (e.g., Will et
al. 1995; de Grijs et al. 2002b; Gouliermis et al. 2006a,b; Kerber et
al. 2006; Da Rio et al. 2009). Recently, the LH 95 IMF obtained by Da
Rio et al. (2009) reached down to $0.31 M_{\odot}$, i.e., much deeper
than achieved previously beyond the Milky Way. Chiosi et al. (2007)
studied the MFs of young SMC clusters down to $0.7 M_{\odot}$. In Liu
et al. (2009), we probed -- for the first time -- the stellar MF in an
extragalactic, low-metallicity environment down to $0.15
M_{\odot}$. In this paper, we applied the method of Liu et al. (2009)
to all of our sample clusters, aimed at assessing the evolution (if
any) of the low-mass MF.

Santiago et al. (2001), de Grijs et al. (2002c), and Kerber et
al. (2006) studied the stellar MFs above $1.0 M_{\odot}$ at different
radii in these same clusters. They found that the slopes at different
radii were significantly different, because significant degrees of
mass segregation affect the PDMF shape. de Grijs et al. (2002b)
studied the NGC 1805 and NGC 1818 MFs above $1.0 M_{\odot}$. They
found that the cluster MFs followed the Salpeter (1955) IMF quite
closely. de Grijs et al. (2002c) studied the LFs of all sample
clusters and concluded that the PDMFs of the clusters in each pair
must be very similar.

The relaxation time at the half-mass radius of compact star clusters
can be written as (Meylan 1987)
\begin{equation}
t_{\rm r,h}=(8.92\times10^{5})\frac{M^{1/2}_{\rm tot}}{\langle m
\rangle}\frac{R^{3/2}_{\rm h}}{\log(0.4M_{\rm tot}/\langle m \rangle)}
\rm yr,
\label{relaxation}
\end{equation}
where $R_{\rm h}$ is the half-mass radius (in pc), $M_{\rm tot}$ the
total cluster mass (in $M_{\odot}$), and $\langle m \rangle$ the
typical mass of a cluster star (in $M_{\odot}$). The timescale on
which a cluster will have lost all traces of its initial conditions is
well represented by its half-mass relaxation time. The dynamical
properties of the Pair I clusters were discussed by de Grijs et
al. (2002b). They (see also de Grijs et al. 2003) computed the
half-mass relaxation time as a function of mass for NGC 1805 and NGC
1818. For stellar masses below $1.0 M_{\odot}$ these were $>300$ and
$>700$ Myr, respectively. This is much longer than the clusters' ages,
which implies that dynamical cluster evolution will not (or
negligibly) affect the MFs below $1.0 M_{\odot}$ at the clusters'
half-mass radii.

Based on the structural parameters of all clusters in Table
\ref{parameters} and their half-mass radii based on the WFPC2 and STIS
observations, we calculated the relaxation times of the other clusters
in pairs II and III for stellar masses below $1.0 M_{\odot}$. These
timescales are $> 2.0$, 0.4, 3.9, and 0.52 Gyr for NGC 1831, NGC 1868,
NGC 2209, and Hodge 14, respectively, although dynamical evolution in
the cluster core may proceed much faster. Elson et al. (1987)
calculated the relaxation time in the core and at the half-mass radius
of NGC 1818 as $\log(t_{\rm r}(0)/{\rm yr}) = 8.2-8.8$ and
$\log(t_{\rm r,h}/{\rm yr}) = 9.0-9.7$ (where the age range signifies
the uncertainties due to the uncertain mass-to-light ratio); for NGC
1831 the equivalent timescales were found to be $\log(t_{\rm
r}(0)/{\rm yr}) = 9.1-9.5$ and $\log(t_{\rm r,h}/{\rm yr}) =
9.6-10.0$. Compared to the clusters' ages (see Table 1), this implies
that both clusters have undergone little to no significant dynamical
evolution overall.

From Figs. \ref{MF.PairI}, \ref{MF.PairII}, and \ref{MF.PairIII} we
conclude that the MF slopes for the entire STIS field of view and
for smaller areas at different radii are identical within the
uncertainties for a given cluster. There may be two reasons for this
behaviour: (i) dynamical cluster evolution does not affect the MFs
severely below $1.0 M_{\odot}$ beyond the core region, although the
relaxation times are shorter than their ages, at least for some
clusters, or (ii) the STIS field is fairly small, which prevents us
from detecting any differences in the MFs below $1.0 M_{\odot}$
beyond the crowded centres.

Because of the uniformity of the MFs derived from the full STIS field
and from smaller areas at different radii, in Fig. \ref{LMCMF} we
compare the MFs of the clusters in each pair using the full STIS
fields only.\footnote{Note that the gap in stellar masses for NGC 2209
between the STIS and WFPC2-based MFs is caused by incompleteness
differences between both sets of observations, and does not
necessarily reflect a true lack of intermediate-mass stars in the
cluster.} In Liu et al. (2009), we adopted both broken power-law and
lognormal distributions to fit the NGC 1818 MFs.  Recent work supports
this method (e.g., Covey et al. 2008; Oliveira et al. 2009). In this
paper, however, it has become clear that not all cluster MFs below
$1.0 M_{\odot}$ show an obvious turnover.  Therefore, we adopted power
laws to fit all cluster MFs below $1.0 M_{\odot}$. We include the
relevant parameters, including those for the standard Kroupa (2001)
IMF, in Table \ref{Slope}. The MF of NGC 1805 shows the same result as
that of NGC 1818 in Liu et al. (2009), i.e., their slopes are
identical to the Kroupa (2001) IMF within the uncertainties, although
the applicable mass ranges extend to higher masses than the relevant
Kroupa (2001) slope (this is likely due to statistical fluctuations;
cf. Liu et al. 2009). NGC 1868 exhibits a broken power-law
distribution similar to the Kroupa (2001) IMF and the complete IMFs of
NGC 1805 and NGC 1818 (de Grijs et al. 2002b; Liu et al. 2009).  The
MF of NGC 1831 does not show a turnover, but its slope in the mass
range from 0.54 to $1.15 M_{\odot}$ is identical to that of the Kroupa
(2001) IMF for masses between 0.5 and $1.0 M_{\odot}$, again within
the uncertainties. The MF slopes of NGC 2209 and Hodge 14 are also
identical to those of the Kroupa (2001) IMF below $1.0 M_{\odot}$.

In this paper, we extended our previous work (Liu et al. 2009) to all
LMC sample clusters. The MFs of the clusters in each pair exhibit
identical slopes, and they are all also identical to the standard
Kroupa (2001) IMF below $1.0 M_{\odot}$, independent of metallicity,
particularly for the clusters in pair I (de Grijs et al. 2002b; Liu et
al. 2009) and NGC 1868 in pair II (this paper). de Grijs et
al. (2002c) studied the LFs of all sample clusters above $1.0
M_{\odot}$. We converted the LFs to MFs based on the Padova isochrones
for the appropriate metallicities and combine both of our results to
obtain complete PDMFs for all sample clusters (see Fig. \ref{LMCMF}
and Table \ref{number.MF}). All MFs are similar to the standard Kroupa
(2001) IMF, and the MFs of NGC 1831, NGC 2209, and Hodge 14 also match
the Chabrier (2003) lognormal distribution, at least qualitatively
although not in detail (as shown by the solid lines in the individual
panels).

\begin{table}
\begin{center}
\begin{minipage}{140mm}
\caption[]{Mass function slopes.}\label{Slope}
\begin{tabular}{ccccc}
  \hline\noalign{\smallskip}
Sample  & \multicolumn{2}{c}{Low mass}  & \multicolumn{2}{c}{High mass} \\
  &  Slope ($\Gamma$) & Mass ($M_{\odot}$) &  Slope ($\Gamma$) & Mass ($M_{\odot}$)\\
\hline\noalign{\smallskip}
Kroupa    & 0.3$\pm$0.5 & 0.08--0.5 & $-$1.3$\pm$0.3 &0.5--1.0\\
(2001)    &             &          & $-$1.3$\pm$0.7 & $\geq$ 1.0\\
NGC 1805  & 0.21$\pm$0.10 & 0.14--0.79 & &\\
NGC 1818  & 0.21$\pm$0.02 & 0.15--0.63 & &\\
NGC 1831  &               & & $-$1.43$\pm$0.11 & 0.54--1.15\\ 
NGC 1868  & 0.28$\pm$0.08 & 0.25--0.63 & $-$1.42$\pm$0.29 & 0.63--0.79\\
NGC 2209  & & & $-$1.23$\pm$0.04 & 0.23--0.60 \\
Hodge14   & & & $-$1.24$\pm$0.16 & 0.46--0.75\\
 \noalign{\smallskip}\hline
\end{tabular}
\end{minipage}
\end{center}

\end{table}

\section{Summary and conclusions}
\label{sect:conclusion}

We extended our pilot study in Liu et al. (2009) and used deep {\sl
HST}/STIS photometry of a carefully selected sample of rich, compact
clusters in the LMC to derive their stellar MFs for masses below $1.0
M_{\odot}$, which we combined with the MFs of de Grijs et al. (2002c)
above $1.0 M_{\odot}$ to obtain complete MFs for all sample
clusters. To our knowledge, together with Liu et al. (2009) this is
the first time that anyone has probed stellar (cluster) MFs to this
depth in an extragalactic, low-metallicity environment. Based on our
STIS observations, the MFs of our compact clusters are all identical
to the standard Kroupa (2001) IMF below $1.0 M_{\odot}$, within the
uncertainties.

The observations were taken beyond the crowded cluster cores and only
included stars at the half-mass radii. In addition, the relaxation
timescales of low-mass stars are much longer than the equivalent
periods for high-mass stars, so that dynamical evolution will not have
affected the younger clusters' stellar mass distributions below $1.0
M_{\odot}$ noticeably. We have therefore provided unprecedented
insights into the IMF in a low-density and low-metallicity
extragalactic environment.

\begin{acknowledgements}

This work was supported by the National Natural Science Foundation of
China under grant No. 10573022 and by the Ministry of Science and
Technology of China under grant No. 2007CB815406. RdG acknowledges
partial financial support from the Royal Society in the form of a
UK--China International Joint Project. We thank Isabelle Baraffe for
advice and discussions in earlier stages of this project. This paper
is based on archival observations with the NASA/ESA {\sl Hubble Space
Telescope}, obtained at the Space Telescope Science Institute, which
is operated by the Association of Universities for Research in
Astronomy, Inc., under NASA contract NAS 5-26555. This research has
made use of NASA's Astrophysics Data System Abstract Service.

\end{acknowledgements}

\end{document}